# Fully automated spectroscopic ellipsometry analyses of crystalline-phase semiconductors based on a new algorithm

Sara Maeda,[1] Kohei Oiwake,[1] Yukinori Nishigaki,[1] Tetsuhiko Miyadera,[2] Masayuki Chikamatsu,[2] Takayuki Nagai,[3] Takuma Aizawa,[4] Kota Hanzawa,[4] Hidenori Hiramatsu,[3,4] Hideo Hosono[3] and Hiroyuki Fujiwara[1,a)]

[1]Department of Electrical, Electronic and Computer Engineering, Gifu University, 1-1 Yanagido, Gifu 501-1193, Japan.
[2]Global Zero Emission Research Center, National Institute of Advanced Industrial Science and Technology (AIST), Central 5, 1-1-1 Higashi, Tsukuba, Ibaraki 305-8568, Japan.
[3]Materials Research Center for Element Strategy, Tokyo Institute of Technology, 4259 Nagatsuta-cho, Midori-ku, Yokohama, Kanagawa 226-8503, Japan.
[4]Laboratory for Materials and Structures, Institute of Innovative Research, Tokyo Institute of Technology, 4259 Nagatsuta-cho, Midori-ku, Yokohama, Kanagawa 226-8503, Japan.

[a)] Author to whom correspondence should be addressed: fujiwara@gifu-u.ac.jp

**Abstract**

One significant drawback of a spectroscopic ellipsometry (SE) technique is its time-consuming and often complicated analysis procedure necessary to assess the optical functions of thin-film and bulk samples. Here, to solve this inherent problem of a traditional SE method, we present a new general way that allows full automation of SE analyses for crystalline-phase semiconductors exhibiting complex absorption features. In particular, we have modified a scheme established in our previous study, which performs a non-linear SE fitting analysis only in a low energy region at the beginning, while the analyzed energy region is gradually expanded toward higher energy by incorporating addition optical transition peaks. In this study, we have further developed a unique analyzing-energy search algorithm, in which a proper analyzing-energy region is determined to incorporate the feature of a new transition peak. In the developed method, a drastic improvement over the previous simple approach has been confirmed for expressing complex dielectric functions consisting of sharp and broad absorption peaks. The proposed method (ΔM method) has been applied successfully to analyze perovskite-based crystalline samples, including hybrid perovskite ($CH_3NH_3PbI_3$) and chalcogenide perovskites ($SrHfS_3$ and $BaZrS_3$). In the automated analyses of these semiconductors, 7 ~ 8 transition peaks are introduced automatically to describe sample dielectric functions, while structural parameters, such as thin-film and roughness thicknesses, are also determined simultaneously. The established method can drastically reduce an analysis time to a level that allows the automatic inspection of daily varying material optical properties and expands the application area of spectroscopic ellipsometry considerably.

# I. INTRODUCTION

For material characterization techniques, it is essential that their measurements provide true physical or chemical properties, desirably in a short period of time with lesser effort. Unfortunately, spectroscopic ellipsometry (SE), employed to characterize sample optical properties (i.e., dielectric function: $\varepsilon = \varepsilon_1 - i\varepsilon_2$) and structures, is an indirect technique whose analysis is generally implemented by using dielectric function and structural models[1-3] and extra efforts are necessary to obtain artifact-free true physical values.[1-4] The difficulties in SE analyses can be viewed as an inverse problem, where the calculation of ellipsometry parameters ($\psi$, $\Delta$) from a dielectric function and an assumed structure is direct and straightforward, while the determination of optical properties and structures from ($\psi$, $\Delta$) is quite difficult, particularly for thin film and multilayer structures.[1-3] As a result, the application of SE is often avoided in material studies and more simple methods with less accuracy, such as transmission and reflection (T/R) methods, have been employed more widely.[5-7]

To solve the significant inverse problem in SE data analyses, we have previously developed a new method, which enables fully automated SE analyses of light absorbing materials in a full measured spectral range without any prior knowledge of analysis parameters.[8] This particular method performs a SE fitting analysis repeatedly, while expanding the analyzed energy region gradually with an addition of a new optical transition peak in the high-energy side whenever the SE fitting error exceeds a critical value. For this method, to find appropriate initial fitting parameters, a zooming-grid-search method has been proposed. By applying the developed scheme, we have previously determined the dielectric functions of $MoO_x$ transparent conductive oxide layers that exhibit band-to-band and deep-level optical transitions.[8] In our earlier attempt, however, the optical functions of amorphous $MoO_x$ layers, which exhibit a broad absorption feature, are analyzed and the applicability for crystalline-phase materials, which typically exhibit sharp-peak-transition features, has not been clear.

In this study, we have developed an improved SE analysis scheme that allows fully automated SE analyses of crystalline-phase semiconductors (hybrid perovskite[2,4,9,10] and chalcogenide perovskite[11-13] materials) exhibiting sharp transition peaks. In particular, in an attempt to establish a general method for the complete automation of SE analyses, we have improved a previous method by adopting a more flexible transition-peak-incorporation algorithm. By using a proposed method, we have demonstrated that the optical properties and structures of general semiconductor materials can be determined automatically in a rather short period of time, while completely avoiding a traditional trial-and-error SE analysis approach.

## II. AUTOMATED SE ANALYSIS METHOD

### A. Overview of the proposed scheme

Figure 1 schematically shows both a basic principle of the fully automated SE analysis developed in our previous study[8] and an improved method proposed in this study. These approaches solve a SE inverse problem by sequential fitting starting from a lower energy region, combined with parameter searches based on an efficient multiple-step grid search (zooming grid search).[8] Specifically, in the automated analysis, the first SE fitting is performed in a limited energy range in a low energy with the highest analyzed range being $E_{HA}$, which is typically set well below the band gap ($E_g$) of semiconductors. For this fitting, a single transition peak, calculated by the Tauc-Lorentz (TL) model[14] in our case, and a conventional optical model (i.e., surface roughness/bulk layer/substrate) can be employed. After performing the SE non-linear regression analysis, the fitting error $\sigma$ is calculated by

$$\sigma = \frac{1}{\sqrt{2L-P}} \{\sum_i^L ([tan\,\psi_{ex}(E_i) - tan\,\psi_{cal}(E_i)]^2 + [cos\,\Delta_{ex}(E_i) - cos\,\Delta_{cal}(E_i)]^2)\}^{1/2}, \quad (1)$$

where $L$ and $P$ show the numbers of measurement points in ellipsometry spectra and analytical parameters, respectively.[1,2] The subscripts of "ex" and "cal" represent experimental and calculated values at the energy of $E_i$. As shown in Fig. 1, if $\sigma$ is lower than a critical value $\sigma_{crit}$, the analyzed energy range is expanded by shifting $E_{HA}$ toward higher energy. The important concept of the proposed scheme is that, when $\sigma > \sigma_{crit}$, an additional transition peak is added to lower $\sigma$ and the SE fitting is continued by incorporating a new TL peak whenever $\sigma$ exceeds $\sigma_{crit}$.

In the example of Fig. 1, $\sigma$ exceeds $\sigma_{crit}$ at 2.0 eV (point A) and the calculated $tan\,\psi$ and $cos\Delta$ spectra indicated by red deviate from the experimental spectra; therefore, a second TL peak is added at this energy. In particular, to maintain high sensitivity for $\sigma$, we calculate $\sigma$ using a fixed high-energy range defined from $E_{HA} - \Delta E_{HA}$ to $E_{HA}$ ($\Delta E_{HA} = 0.5$ eV in our study). In other words, $L$ is constant in Eq. (1). In Fig. 1, this $\sigma$ is calculated from 1.5 ~ 2.0 eV when the second TL peak is added and this energy region is indicated by a green area with the energy width of $w_1$. This procedure established in our previous study,[8] however, has one drawback; $\sigma$ is rather insensitive to the actual shape of the arrow), whereas in our previous method[8] the upper limit of the analyzed range is fixed by the condition of $\sigma > \sigma_{crit}$ (green arrow). The green region with an energy width of $w_1$ shows the SE fitting region in the previous scheme, whereas the orange region with a width of $w_2$ indicates the SE fitting region with the optimized $\Delta M$ of 0.4 eV. The SE fitting results for each $\sigma$ point of A-D are also shown.

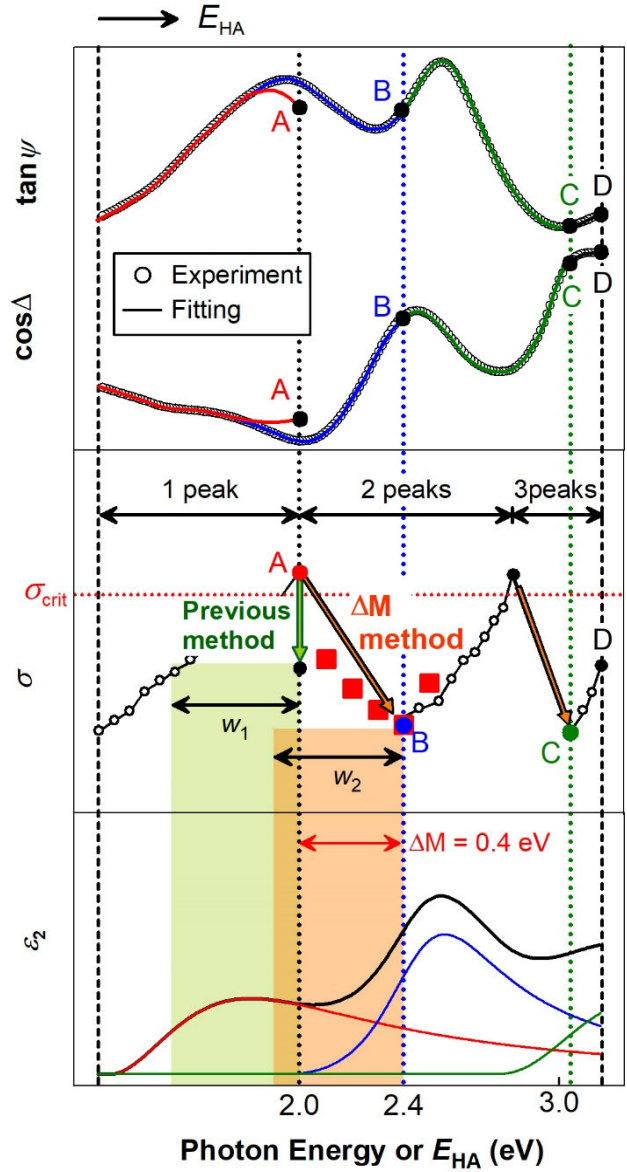

**FIG. 1.** Procedure of a fully automated SE analysis proposed in this study. In the established method, experimental $\tan\psi$ and $\cos\Delta$ ellipsometry spectra (open circles) are analyzed sequentially from a lower energy side and the region of the SE fitting (solid lines) is expanded gradually toward the higher energy. When the SE fitting error $\sigma$ obtained in a selected energy region becomes higher than a critical value ($\sigma_{crit}$), a new transition peak expressed by the Tauc-Lorentz (TL) model is incorporated. In this example, only one TL peak is used when the highest analyzed energy ($E_{HA}$) is $\leq 2.0$ eV, whereas two and three TL peaks are employed in the fitting analyses in the regions $2.0 \leq E_{HA} \leq 2.8$ eV and $E_{HA} \geq 2.8$ eV, respectively. For the absorption peak incorporation, the $\Delta M$ method searches the optimum analyzing-energy region with a small incremental energy of $\Delta M$ (indicated by red squares) and the optimum $\Delta M$ value is determined (orange

--------------------------------------------------------------------------------

second TL peak shown by a blue line in Fig. 1, because $\sigma$ is calculated in the region where the $\varepsilon_2$ contribution of the second TL peak is negligible.

In this study, the above method has been improved by sequentially shifting the $\sigma$ calculation range with a small additional energy of $\Delta M$. Specifically, when the second TL peak is added, $\sigma$ is calculated by systematically shifting the analyzing energy region from $E_{HA} - \Delta E_{HA} + \Delta M$ to $E_{HA} + \Delta M$. In the case of Fig. 1, for example, $\Delta M$ is increased from 0.1 to 0.5 eV and the $\sigma$ values for each $\Delta M$ (a total of five points) are calculated, as depicted by red squares. In this second step, we select $\Delta M$ which provides the minimum $\sigma$ as the optimized analyzing energy region. In Fig. 1, $\Delta M = 0.4$ eV provides the lowest $\sigma$, leading to the best fitting up to $E_{HA} + \Delta M = 2.4$ eV (indicated by blue tan$\psi$ and cos$\Delta$ spectra). This procedure completes the addition of the second TL peak. It can be seen from Fig. 1 that this method ($\Delta M$ method) allows the $\sigma$ evaluation in a higher energy range, as indicated by an orange region of $w_2$, and indeed this expanded region shows a more significant overlap with the second $\varepsilon_2$ peak. In other words, although the procedure of $\Delta M$ is quite simple, the $\Delta M$ method enables us to include the influence of a new incorporating transition peak more precisely. After the $\Delta M$ energy is selected for the second peak, the sequential SE fitting with an expansion of $E_{HA}$ is started again and a third peak is introduced at the energy point C. Accordingly, a whole process is repeated by adding a new TL peak whenever $\sigma > \sigma_{crit}$ until $E_{HA}$ reaches the maximum measured energy. It should be emphasized that the $\Delta M$ method becomes particularly important when analyzed samples are crystalline-phase materials, which generally show sharp absorption features.

**B. Procedure of the proposed method**

Figure 2 shows the flow chart of the developed automated SE analysis. The proposed method essentially follows the scheme established in our previous study,[8] but the modifications incorporated particularly in the ΔM method are shown by red in Fig. 2. In the flow chart, after setting initial parameters ($\sigma_{crit}$ and $E_{HA}$), a SE fitting analysis is performed. When $E_{HA} < E_g$, there is no significant light absorption (i.e., $\varepsilon_2 \sim 0$) and we can employ a simple optical function model in this region. In this study, the analysis for the region of $\varepsilon_2 \sim 0$ was first implemented using the Cauchy model [$n(\lambda) = A + B/\lambda^2 + C/\lambda^4$] assuming a thin film structure described by a surface roughness layer thickness ($d_s$) and a bulk layer thickness ($d_b$). Thus, by performing the SE fitting analysis using five parameters, the fitting error $\sigma$ is calculated by Eq. (1). The analyzed energy region ($E_{HA}$) is then expanded until we observe $\sigma > \sigma_{crit}$. At $E_{HA}$ of $\sigma > \sigma_{crit}$, we perform the zooming grid search[8] to find the first set of initial guess parameters for the first TL peak and structural parameters (i.e., $d_s$ and $d_b$) and, by applying these guess parameters, a non-linear SE fitting analysis (all parameter fitting) is carried out. After this step, the analyzed energy region is expanded by $\Delta E$ (i.e., $E_{HA} = E_{HA,old} + \Delta E$ and $\Delta E = 0.1$ eV in this study). As established earlier,[8] after the $E_{HA}$ expansion, we perform a restricted SE fitting, in which only the transition peak having the highest peak energy is fitted. If $\sigma > \sigma_{crit}$, a new transition peak is incorporated based on the ΔM method. When the new peak is included, the zooming grid search and all parameter fitting are also performed for different ΔM values. When the introduction of a new peak does not improve the fitting (i.e., when $\sigma > \sigma_{crit}$ even after the peak addition), $\sigma_{crit}$ is increased by $\Delta\sigma$ ($\sigma_{crit} = \sigma_{crit,old} + \Delta\sigma$) and, in our case, $\Delta\sigma = 1 \times 10^{-3}$. Accordingly, $\sigma_{crit}$ is not a predetermined fixed value but is a variable which increases according to the situation of the SE fitting. In contrast, if $\sigma$ is lowered successfully by the peak addition (i.e., in case $\sigma < \sigma_{crit}$), $E_{HA}$ is expanded again and the whole procedure is repeated until the analyzed energy region reaches the maximum energy of measured experimental spectra ($E_{max}$). It should be emphasized that the final SE results are obtained by all parameter fitting where all the parameter values for the TL transition peaks and structure are varied to obtain the best fitting values. The detailed procedure can be found in our earlier study[8] and the important improvements in the present study are (i) the incorporation of the Cauchy model in the analysis of $\varepsilon_2 \sim 0$ and (ii) the introduction of the ΔM method for the peak incorporation. We also mention that, when we observe an analysis error in the SE fitting analysis, the error is handled essentially in the same way as previously.[8]

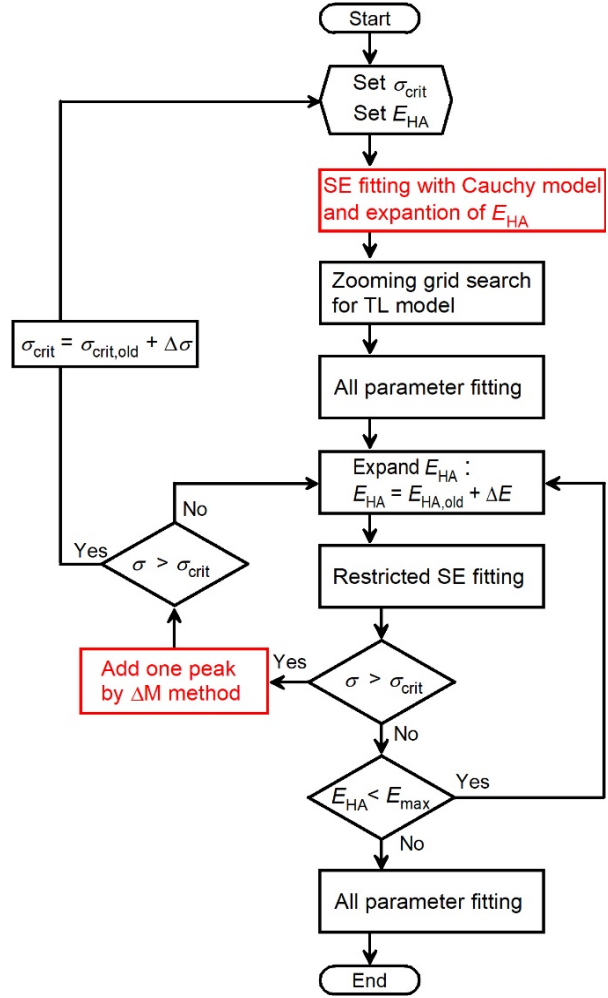

**FIG. 2.** Flow chart of the fully automated SE analysis based on the ΔM method. The modifications over our previous method are (i) the incorporation of the Cauchy model in the analysis region of $\varepsilon_2 \sim 0$ and (ii) the introduction of the ΔM method for the peak addition, as indicated by red squares in the figure. The detail of the other processes can be found in Ref. 8. Note that the developed scheme has only two initial setting parameters of $\sigma_{crit}$ and $E_{HA}$ ($\sigma_{crit} = 1 \times 10^{-3}$ and $E_{HA} = 1.3$ eV in the first fitting). The $E_{HA}$ is increased sequentially with an energy step of $\Delta E$ (0.1 eV in this study). If the introduction of an additional TL peak does not improve the fitting quality (i.e., $\sigma$), $\sigma_{crit}$ is increased by a step of $\Delta \sigma$ ($1 \times 10^{-3}$ in this study). When $E_{HA}$ reaches $E_{max}$, all parameter fitting is performed to obtain the final fitting result. The zooming grid search is a multiple-step grid search method developed earlier,[8] in which the first grid search is carried out using a coarse mesh and the second and third grid searches are further performed using finer grids by zooming the grid around the guess values obtained in the precedent grid search.

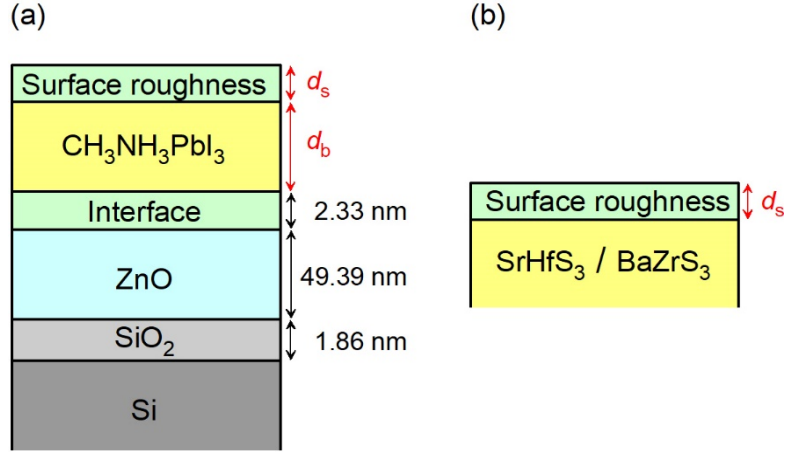

**FIG. 3.** Optical models and sample structures of (a) the $CH_3NH_3PbI_3$ hybrid perovskite and (b) the $SrHfS_3$ and $BaZrS_3$ chalcogenide perovskite samples. The hybrid perovskite has a thin film structure,[10] whereas the chalcogenide perovskites are bulk samples.[13] The thicknesses of the underlying layers, determined by separate SE analyses, are also indicated for (a).

### III. SEMICONDUCTOR ANALYSES

For the automated SE analyses based on the newly developed ΔM method, the ($\psi$, $\Delta$) spectra obtained from a hybrid perovskite ($CH_3NH_3PbI_3$) thin film[10] and chalcogenide perovskite ($SrHfS_3$ and $BaZrS_3$) bulk samples[13] were applied. These materials are direct transition solar-cell materials exhibiting strong light absorption near the band edges.[4,10-13] In the case of the hybrid perovskite, to avoid intense degradation in air,[4,15] the ellipsometry spectra were measured without exposing the sample to air at all.[10] Figure 3 shows the optical models (sample structures) of (a) the hybrid perovskite and (b) the chalcogenide perovskite samples. The $CH_3NH_3PbI_3$ layer with a thickness of ~ 45 nm was formed on a ZnO-coated $SiO_2$/Si substrate by a laser evaporation technique[10] and the structure (layer thickness) of the underlying substrate is also indicated in Fig. 3(a). The ZnO layer has been provided to improve the homogeneity of the depositing perovskite layer. The $CH_3NH_3PbI_3$/ZnO interface layer thickness corresponds to the surface roughness layer thickness of the ZnO layer obtained in the SE analysis of the ZnO/$SiO_2$/Si substrate structure.[2,16] In the SE analysis of Fig. 3(a), the dielectric function of the surface roughness of $CH_3NH_3PbI_3$ and the interface layer of $CH_3NH_3PbI_3$/ZnO were obtained by applying Bruggeman effective medium approximation (EMA)[1,2,17] assuming a 50/50 vol.% of void/$CH_3NH_3PbI_3$ and $CH_3NH_3PbI_3$/ZnO components, respectively.

The chalcogenide perovskite bulk polycrystals were made by a solid-state reaction method with the size of the synthesized pellets being 6 - 7 mm in diameter and 1 mm in thickness.[13] After the syntheses, the bulk samples were polished to obtain the optical-grade mirror surface for the SE measurements. However, due to the imperfection of the mechanical polishing, the surface roughness layer was provided in the optical model, as shown in Fig. 3(b). The dielectric function of the roughness layer was obtained from the Bruggeman EMA.

## IV. RESULTS AND DISCUSSION

### A. SE analysis of the hybrid perovskite

Figure 4 shows the results of the automated SE analysis performed for the hybrid perovskite thin film, obtained from the ΔM method with the expansion of the $E_{HA}$ range up to (a) 1.6 eV, (b) 2.5 eV, (c) 4.0 eV, and (d) 5.1 eV. The analyzed region of each analysis is indicated by pale yellow. The open circles show the experimental $\tan\psi$ and $\cos\Delta$ spectra and the solid lines indicate the calculated spectra. The corresponding $\varepsilon_2$ spectra obtained assuming the TL transitions are also indicated, whereas $N$ in each panel denotes the number of the TL transition peaks used in each analysis.

In Fig. 4(a), only one TL peak is used in the analysis and the good fitting is observed only in a limited analyzed region of 0.75 ~ 1.6 eV, while the calculated $\tan\psi$ and $\cos\Delta$ spectra deviate significantly outside the analyzed region due to the simple assumption of $N = 1$. With the expansion of $E_{HA}$ [i.e., (a) → (d) in Fig. 4], new absorption peaks are introduced automatically and the number of $N$ increases. The result of Fig. 4 demonstrates that the ΔM method provides the excellent fitting for all the analyzed regions, even though the hybrid perovskite exhibits sharp absorption features above $E_g$ = 1.61 eV.[10] It can also be confirmed that some TL peaks are located outside the analysis range but these transition peaks are modified properly as $E_{HA}$ increases.

Figure 5 further summarizes the detail of the automated SE analysis of the hybrid perovskite as a function of $E_{HA}$ and photon energy: (a) $\sigma$, (b) $\tan\psi$, (c) $\cos\Delta$, (d) $\varepsilon_2$, and (e) $d_b$ and $d_s$ values. In Fig. 5, the result up to $E_{HA}$ = 1.6 eV was obtained using the Cauchy model, while the analysis result at $E_{HA} \geq$ 1.6 eV was determined assuming the TL optical transitions. The variation of $\sigma$ with $E_{HA}$ is shown in Fig. 5(a) and the ΔM analysis of the hybrid perovskite shows that the appropriate $\sigma_{crit}$ is $2.0 \times 10^{-3}$. The arrows in top image indicate $E_{HA}$ at which a new TL peak is incorporated and the closed circles in Fig. 5(a) represent the $\sigma$ values obtained after the peak addition based on the ΔM approach. As a result, the dielectric function of $CH_3NH_3PbI_3$ is expressed by a total of eight peaks. In Figs. 5(b)-(d), the results for the SE fittings and the resulting $\varepsilon_2$ spectra are shown. The $\sigma$ calculated for the whole spectral region of Figs. 5(b) and 5(c) is $\sigma_{total} = 2.06 \times 10^{-3}$ and thus the fitting error is sufficiently low.

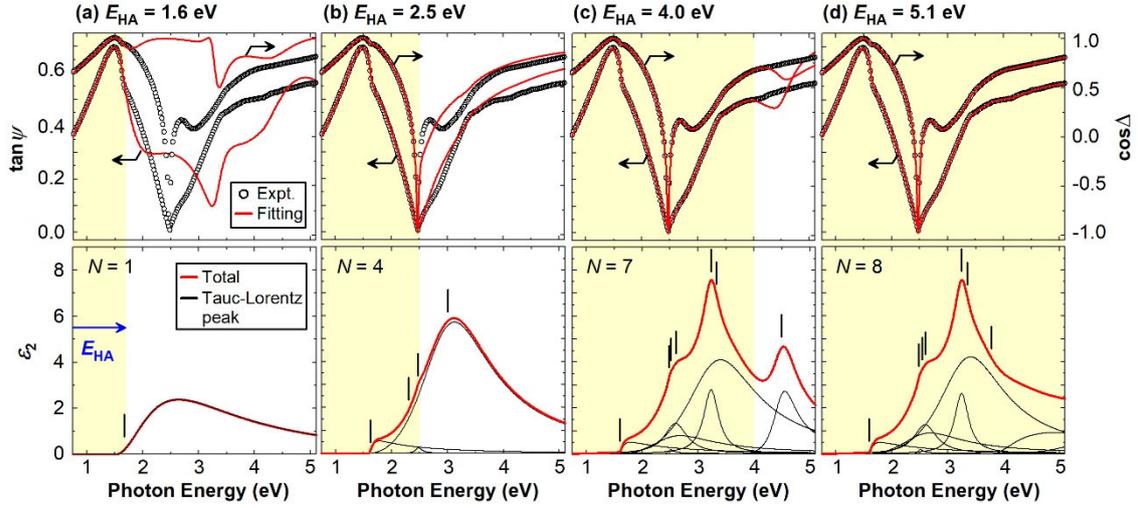

**FIG. 4.** Automated SE analyses of the hybrid perovskite thin film by the ΔM method with the expansion of the $E_{HA}$ range up to (a) 1.6 eV, (b) 2.5 eV, (c) 4.0 eV, and (d) 5.1 eV. The analyzed region of each analysis is indicated by pale yellow. The open circles show the experimental tan$\psi$ and cos$\Delta$ spectra, whereas the solid lines indicate the calculated spectra. The corresponding $\varepsilon_2$ spectra obtained assuming the TL transitions are also indicated, whereas N in each panel indicates the number of the TL transition peaks used in each analysis. The bars in the $\varepsilon_2$ spectra indicate the energy positions of the $E_0$ transition energy in the TL model.

In Fig. 5(e), the changes of $d_b$ and $d_s$ are correlated and indeed the total thickness defined by $d_{total} = d_b + 0.5 d_s$ is almost constant within 1.5 nm. Here, the coefficient of 0.5 for $d_s$ corresponds to the void volume fraction assumed for the surface roughness. The final SE result corresponds to the ones obtained at $E_{HA}$ = 5.1 eV and the structural parameters estimated at this energy ($d_b$ = 44.98 nm and $d_s$ = 1.34 nm) are the representative values of the thin film structure.

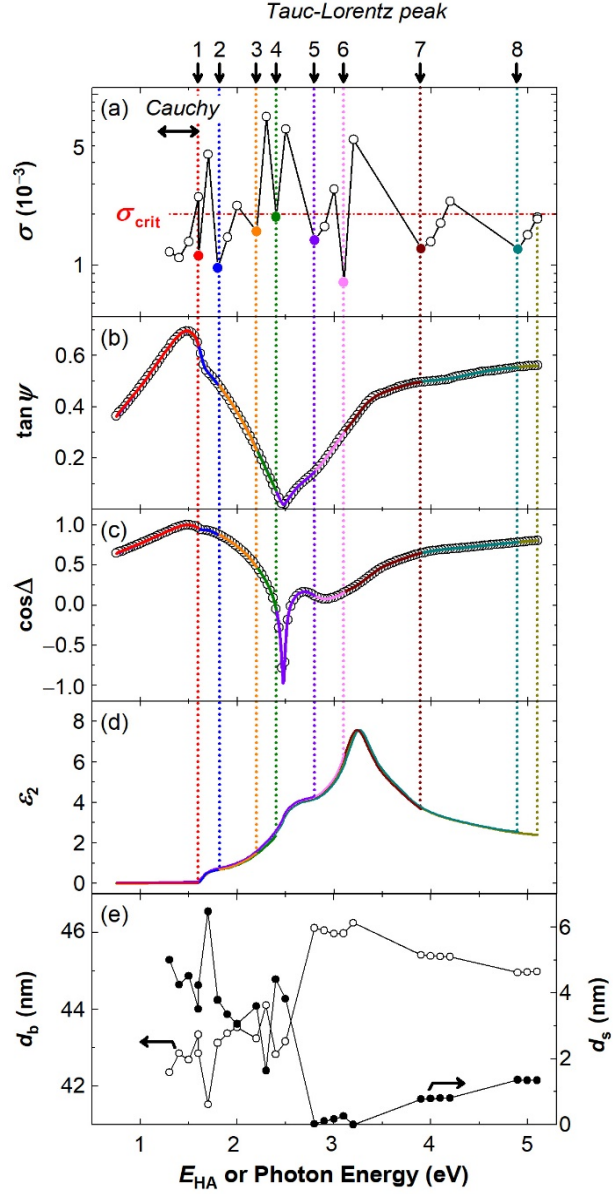

**FIG. 5.** Results of the fully automated SE analysis performed for the hybrid perovskite as a function of $E_{HA}$ and photon energy: (a) $\sigma$, (b) $\tan\psi$, (c) $\cos\Delta$, (d) $\varepsilon_2$, and (e) $d_b$ and $d_s$ values. The arrows in top image indicate the $E_{HA}$ at which additional TL peaks are introduced. The results of $E_{HA} \leq 1.6$ eV were obtained using the Cauchy model, while the analysis results at $E_{HA} \geq 1.6$ eV were determined assuming the TL optical transitions. The open circles in (a) indicate $\sigma$ at each $E_{HA}$, whereas the closed circles indicate $\sigma$ obtained after the energy-search and the following SE fitting by the $\Delta M$ method. The appropriate $\sigma_{crit}$ of this sample is $2.0 \times 10^{-3}$.

In the automated analysis, a total of eight peaks are introduced to express the dielectric function of the hybrid perovskite. Each TL transition peak is expressed by four parameters[14] and, therefore, the $\varepsilon_2$ spectrum is described by a total of 32 parameters in the case of $CH_3NH_3PbI_3$. With the addition of an $\varepsilon_1$ offset value[14] and the structural parameters, the ($\psi, \Delta$) spectra are expressed by 35 independent variables. Accordingly, our $\Delta M$ method solves the complicated inverse problem based on a unique peak incorporation scheme by employing an extensive number of parameters, which are determined automatically by the developed algorithm shown in Fig. 2. Moreover, in the $CH_3NH_3PbI_3$ analysis, the zooming grid search has been implemented repeatedly with a total of over $10^6$ mesh points and the number of non-liner SE fittings performed in the whole analysis is ~100. In other words, the accurate SE analysis based on the $\Delta M$ method is realized by the vast number of guess-parameter combinations and the following optimizations.

For the hybrid perovskite dielectric function, the contribution of each transition peak determined in Fig. 5 is shown in the supplementary material Fig. 1. The obtained $\varepsilon_2$ peak contributions are slightly different from those determined from the manual $\varepsilon_2$ peak deconvolution analysis reported earlier[3] (supplementary material Fig. 1), even though the total $\varepsilon_2$ spectra are almost identical.

Figure 6 shows the $\varepsilon_2$ spectrum near and below $E_g$ of $CH_3NH_3PbI_3$, obtained from the $\Delta M$ method (blue line). In this spectrum, the small $\varepsilon_2$ value can be confirmed even below $E_g$ of 1.61 eV, which is an artifact generated by the automated analysis. Specifically, when the number of the TL transition peaks is increased, the $\Delta M$ method tries to lower $\sigma$ by incorporating the small $\varepsilon_2$ contribution. In the analysis of Fig. 5, when $N$ = 5, the $E_g$ of the newly added peak becomes much smaller than the $E_g$ of the material and this nominal absorption lowers $d_s$ to ~ 0 nm at $E_{HA}$ = 2.8 eV. As confirmed previously,[2,18] $d_s$ is highly correlated with the $\varepsilon_2$ component and the phase lag (i.e., $\Delta$ value) within the roughness layer can be misinterpreted as the light absorption in the material. We found that the $\Delta M$ method tends to perform the "overfitting", where the algorithm tries to obtain the better fitting particularly for the low-energy $\Delta$ spectrum by increasing $\varepsilon_2$ slightly, even though there is actually no observable light absorption in a material. Unfortunately, at this stage, the effective method for preventing this overfitting has not been found.

In this study, therefore, we have performed an extra roughness correction (ERC)[2,18] for the $\Delta M$-derived $\varepsilon_2$ spectrum to obtain the artifact-free dielectric function. As established previously,[2,18] the nominal $\varepsilon_2$ tail absorption originates from the underestimated $d_s$ contribution. Thus, if this $d_s$ factor is corrected, we can obtain the correct optical function. In other words, the blue $\varepsilon_2$ spectrum in Fig. 6 is the pseudo-dielectric function containing the influence of the roughness and the roughness thickness needs to be corrected by removing the extra roughness. This ERC can be performed quite easily by removing $d_s$ in a roughness/bulk optical model until the artificial band-edge absorption is removed completely and $\varepsilon_2$ becomes zero below $E_g$.

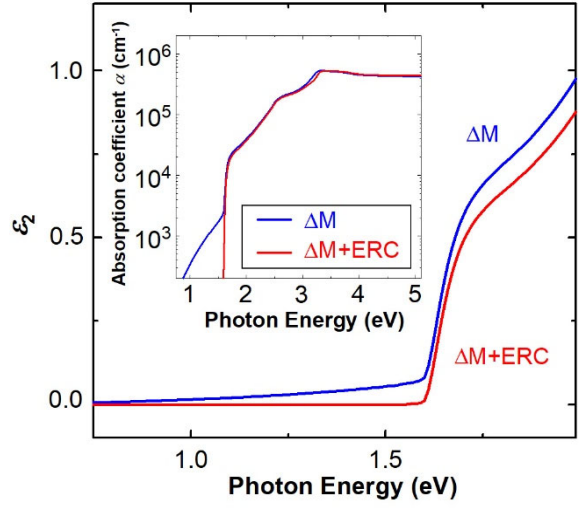

**FIG. 6.** $\varepsilon_2$ spectra near and below $E_g$ of CH$_3$NH$_3$PbI$_3$, obtained before (blue line) and after (red line) the extra roughness correction (ERC).[18] The original $\varepsilon_2$ spectrum obtained from the ΔM method (blue line) corresponds to that shown in Fig. 5(d). The small $\varepsilon_2$ contribution observed below $E_g$ = 1.61 eV is an artifact generated by the overfitting in the automated analysis. This artifact can be removed completely by the ERC. The inset shows the corresponding absorption coefficient spectra obtained before (blue line) and after (red line) the ERC.

The red $\varepsilon_2$ spectrum in Fig. 6 represents the $\varepsilon_2$ spectrum obtained after the ERC. It should be emphasized that the overall $\varepsilon_2$ magnitude does not vary by the ERC and the ERC corrects the $\varepsilon_2$ value only slightly (supplementary material Fig. 2). For the perovskite, the corrected roughness value is $\Delta d_s$ = 0.75 nm; thus a final $d_s$ of the perovskite layer is $d_s$ = 1.34 nm (Fig. 5) + 0.75 nm (ERC) = 2.09 nm. The atomic force microscopy of the sample provides the root-mean-square roughness of $d_{rms}$ = 4.6 nm,[10] which reasonably agrees with the SE result.[2,17] The inset of Fig. 6 shows the corresponding absorption coefficient ($\alpha$) spectra obtained before (blue) and after (red) the ERC. It can be confirmed that the ERC procedure changes $\alpha$ predominantly in the below-band-gap absorption. Many studies have confirmed that $CH_3NH_3PbI_3$ does not show tail absorption and exhibits a very sharp absorption edge with the Urbach energy of 14 meV.[4,6,10]

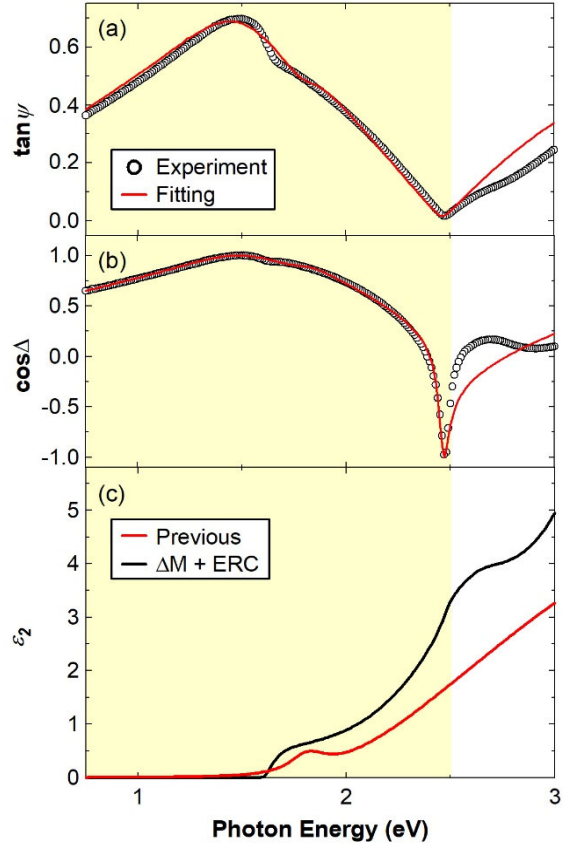

**FIG. 7.** SE results obtained by applying our previous method developed by Oiwake et al.[8] (red line) and a newly developed ΔM + ERC method (black line): (a) tan$\psi$, (b) cos$\Delta$, and (c) $\varepsilon_2$ spectra. The region of pale yellow ($E_{HA}$ = 2.5 eV) indicates the SE analysis range for the previous method.

We further find that the fully automated analysis of the $CH_3NH_3PbI_3$ thin film cannot be performed by the method established in our previous study[8] and it is vital to apply the ΔM method to perform the accurate analysis. Figure 7 summarizes the SE results obtained by applying our previous method with $E_{HA}$ = 2.5 eV (red line): (a) $\tan\psi$, (b) $\cos\Delta$, and (c) $\varepsilon_2$ spectrum. In Fig. 7(c), the correct $\varepsilon_2$ spectrum obtained from the ΔM method with the ERC (black line) is also shown. When our previous method is adopted, $\sigma$ increases sharply with increasing $E_{HA}$ and it becomes quite difficult to obtain a satisfactory fitting at high energies ($E_{HA}$ > 2.5 eV). In particular, the $\varepsilon_2$ spectrum obtained from our previous approach is quite broad, whose shape is quite different from the accurate spectrum. Moreover, a fine absorption feature observed in the $\tan\psi$ and $\cos\Delta$ spectra at ~1.6 eV, originating from the band-gap transition, cannot be fitted well using the previous method. Accordingly, the ΔM method is a quite important improvement over our previous method.

**B. SE analyses of the chalcogenide perovskites**

In order to confirm the validity of the ΔM method, additional two samples ($SrHfS_3$ and $BaZrS_3$) were analyzed based on the ΔM method. Figure 8 summarizes the results of the fully automated ΔM analysis for the $SrHfS_3$ bulk sample: (a) $\sigma$, (b) $\tan\psi$, (c) $\cos\Delta$, (d) $\varepsilon_2$, and (e) $d_s$. The overall analysis procedure of $SrHfS_3$ is exactly the same as that of $CH_3NH_3PbI_3$, although the optical model is different, as shown in Fig. 3(b). The ΔM analysis of the $SrHfS_3$ chalcogenide perovskite shows that the appropriate $\sigma_{crit}$ is $1.0 \times 10^{-2}$ and the $SrHfS_3$ $\varepsilon_2$ spectrum has been determined successfully up to 6.0 eV by incorporating a total of seven TL peaks. The SE fitting analysis was switched from the Cauchy to the TL model at $E_{HA}$ = 2.4 eV, which approximately corresponds to the $\varepsilon_2$ onset of $SrHfS_3$, and the excellent fitting has been confirmed for the $\tan\psi$ and $\cos\Delta$ spectra in the whole analyzed region with the $\sigma$ calculated for the whole spectral region of Figs. 8(b) and 8(c) being $\sigma_{total} = 4.75 \times 10^{-3}$. The $d_s$ value changes largely when $E_{HA}$ is small but converges to $d_s$ = 3.51 nm at high energies.

Figure 9 shows the (a) $\varepsilon_1$, (b) $\varepsilon_2$, and (c) enlarged $\varepsilon_2$ spectra of the $SrHfS_3$ sample, obtained from the automated analysis of Fig. 8. The $SrHfS_3$ exhibits rather broad absorption peaks at 3.6 and 4.3 eV. However, the absorption tail of the second and third TL peaks extends toward lower energy below $E_g$ = 2.41 eV and thus the problem of the "overfitting" also occurs in the automated analysis of the $SrHfS_3$. As shown in Fig. 9(c), the ERC can remove the artificial tail absorption using the roughness correction value of 2.0 nm. We also mention that, when our previous automated analysis method was applied to the $SrHfS_3$ analysis, the satisfactory dielectric function was obtained only up to 3.2 eV. Thus, our previous method has a limited applicability.

We have performed the automated SE analysis for the other chalcogenide perovskite

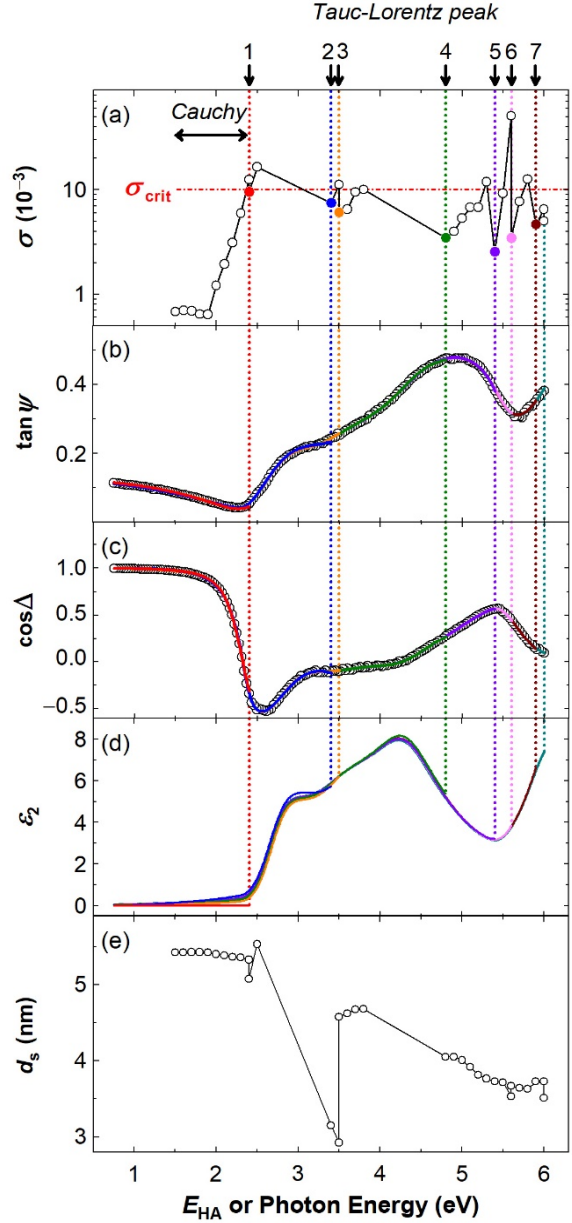

**FIG. 8.** SE results obtained from the fully automated ΔM analysis for the SrHfS$_3$ bulk sample: (a) $\sigma$, (b) tan$\psi$, (c) cosΔ, (d) $\varepsilon_2$, and (e) $d_s$. The arrows in top image indicate the $E_{HA}$ at which additional TL peaks are introduced. The results of $E_{HA} \leq 2.4$ eV were obtained using the Cauchy model, while the analysis results at $E_{HA} \geq 2.4$ eV were determined assuming the TL optical transitions. The open circles in (a) indicate $\sigma$ at each $E_{HA}$, whereas the closed circles indicate $\sigma$ obtained after the analyzing-energy search by the ΔM method and the following SE fitting. The appropriate $\sigma_{crit}$ of this sample is $1.0 \times 10^{-2}$.

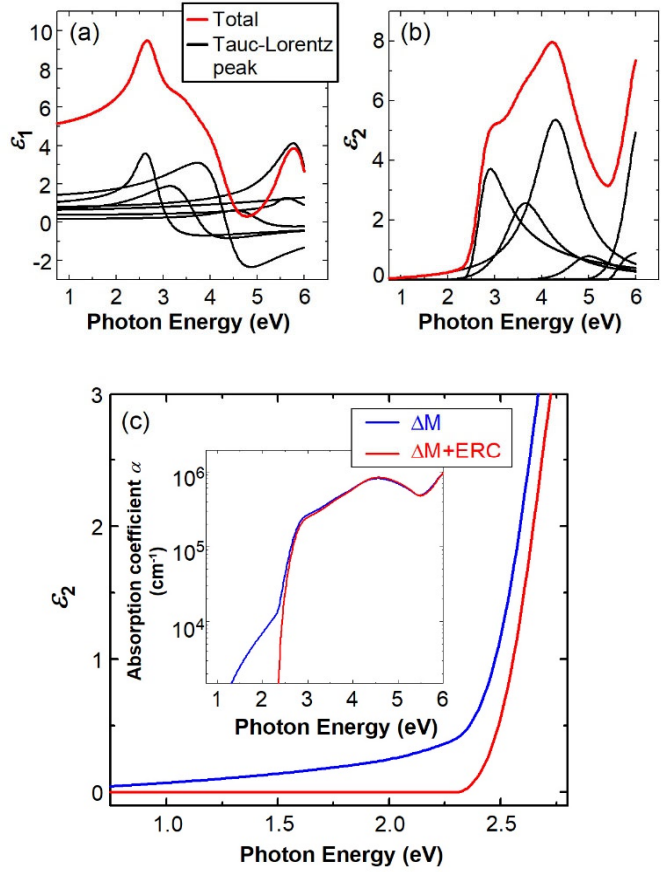

**FIG. 9.** (a) $\varepsilon_1$, (b) $\varepsilon_2$, and (c) enlarged $\varepsilon_2$ spectra of the SrHfS$_3$ sample, obtained from the automated analysis of Fig. 8. In (a) and (b), the contributions of each TL transition peak are shown by black, whereas the red lines indicate the total $\varepsilon_1$ and $\varepsilon_2$ spectra. The red $\varepsilon_2$ spectrum in (b) corresponds to that shown in Figure 8(d). The tail absorption observed in (b) and (c) is an artifact caused by the "overfitting". In the inset of (c), the $\alpha$ spectra obtained from the ΔM method (blue) and ΔM + ERC method (red) are shown.

(BaZrS$_3$). The overall analysis results obtained for the BaZrS$_3$ bulk sample (supplementary material Figures 3 and 4) are essentially quite similar to those obtained for the SrHfS$_3$, as the dielectric functions of SrHfS$_3$ and BaZrS$_3$ are similar, even though the $E_g$ of BaZrS$_3$ (1.94 eV) is slightly lower than that of SrHfS$_3$ (2.41 eV). In the ΔM analysis of the BaZrS$_3$ sample, we also confirmed the "overfitting" problem, which can be corrected properly by the ERC with the roughness correction value of 2.7 nm for this case. The number of the total TL peaks used for the automated analysis of the $BaZrS_3$ is seven, which is identical to that of the $SrHfS_3$. Accordingly, the variation of the dielectric function with atomic species can be determined systematically by applying the same automated analysis method developed in this study.

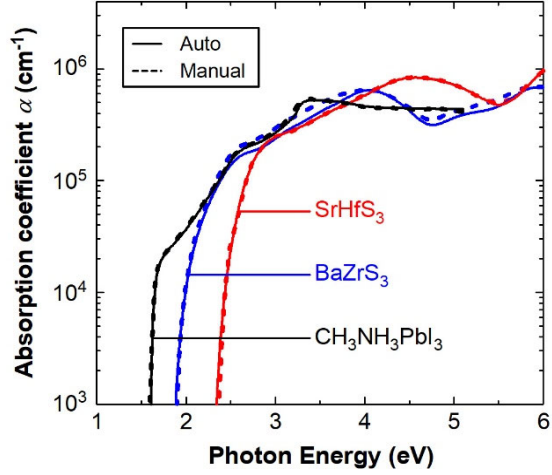

**FIG. 10.** $\alpha$ spectra of $CH_3NH_3PbI_3$, $SrHfS_3$, and $BaZrS_3$ perovskites obtained from the automated SE analyses based on the $\Delta M$ method with the ERC (solid lines) and the manual SE analyses reported earlier[10,13] (dotted lines). The manual analysis results have been obtained by SE analyses with mathematical inversion, followed by the dielectric function parameterization assuming the TL transition peaks.

### C. Comparison with the manual SE analyses

Figure 10 compares the $\alpha$ spectra of $CH_3NH_3PbI_3$, $SrHfS_3$, and $BaZrS_3$ perovskites obtained from the above automated SE analyses based on the $\Delta M$ + ERC method (solid lines) with those determined from the manual SE analyses reported earlier[10,13] (dotted lines). The manual analysis results have been obtained by SE analyses with mathematical inversion, followed by the dielectric function parameterization assuming the TL transition peaks.

It can be seen that the optical functions obtained from the $\Delta M$ + ERC method are almost identical to those obtained from the manual analyses, which justifies the validity of the automated SE analyses. From the automated SE analysis method, the structural parameters (film thickness, roughness), in addition to the dielectric function modeling parameters, can be obtained automatically by applying the exact same algorithm. It should be emphasized that SE combined with the ΔM + ERC method eliminates the significant drawback of a time-consuming and complicated manual SE analysis, allowing the optical analysis of various materials in an easy manner with an accuracy far exceeding a traditional transmission/reflection (T/R) technique.

## VI. CONCLUSION

We have developed a new algorithm that can perform fully automated SE analyses of crystalline semiconductor materials exhibiting sharp absorption features with narrow transition peaks. The established method allows the complete determination of material dielectric functions in a full measured spectral range and sample structures without any prior knowledge of material optical properties and initial guess parameters. In our method, the initial SE fitting is restricted in a low energy region and the analyzed energy region is gradually expanded toward higher energy while introducing an additional transition peak whenever the fitting quality exceeds a critical value until the dielectric function is determined in a full spectral range. The important feature of the improved method (ΔM method) is that the systematic energy-search algorithm is incorporated for the SE analyzed region when a new optical transition peak is introduced. This approach allows the incorporation of the new absorption peak with proper transition energies, enabling us to improve SE fitting quality effectively. In particular, a reliable algorithm has been established and the dielectric functions of direct-transition perovskite semiconductors ($CH_3NH_3PbI_3$, $SrHfS_3$, and $BaZrS_3$) have been analyzed successfully. In these crystalline-phase analyses, 7 ~ 8 TL transition peaks are introduced automatically, providing sufficiently good SE fittings in the whole analyzed region. From the fitting analyses, the structural parameters of the samples have been extracted simultaneously. The established method is quite general and further modification appears to be unnecessary.

Nevertheless, all the obtained SE results show that the ΔM method tends to lower the SE fitting error by providing a fictious tail absorption in energies well below $E_g$ of semiconductors. Quite fortunately, this "overfitting" problem can be corrected by performing the extra roughness correction established previously. When the ΔM method is combined with the extra roughness correction, the optical functions exhibit remarkable agreement with those determined by a trial-and-error approach. The developed method is expected to contribute to open a novel way for SE as a more general characterization tool that allows the accurate determination of material structures and optical properties in a short time.

**Supplementary Material**

See the supplementary material for the contributions of the TL transition peaks of $CH_3NH_3PbI_3$ (Fig.1), the $\varepsilon_1$ and $\varepsilon_2$ spectra before and after the ERC of the $CH_3NH_3PbI_3$ (Fig. 2), the SE analysis results (Fig. 3) and dielectric function (Fig. 4) obtained for the $BaZrS_3$ bulk sample.

**Data availability**

The data that support the findings of this study are available from the corresponding author upon reasonable request.

Tampo, M. Chikamatsu, and H. Shibata, Appl. Surf. Sci., **421**, 276-282, (2017)